\documentclass[12pt]{article}

\usepackage{psfig}

\newlength{\figwidth}
\newcommand{\ealine}[3]
{$\displaystyle #1$ & $\displaystyle #2$ & $\displaystyle #3$}

\newcommand{\eaABCbe}[1] { 
\refstepcounter{equation}\label{e:#1}
\begin{center}
\begin{tabular}{rcl}
}

\newcommand{\eaABen}[2]{
\end{tabular}
\\[#1] \makebox[\textwidth][r]{(\theequation a)}
\\[#2] \makebox[\textwidth][r]{(\theequation b)}
\end{center}
}
\newcommand{\eaABCen}[3]{
\end{tabular}
\\[#1] \makebox[\textwidth][r]{(\theequation a)}
\\[#2] \makebox[\textwidth][r]{(\theequation b)}
\\[#3] \makebox[\textwidth][r]{(\theequation c)}
\end{center}
}

\newcommand{\eabe} {\begin{eqnarray}}
\newcommand{\eaen} {\end{eqnarray}}
\newcommand{\eqbe} {\begin{equation}}
\newcommand{\eqen} {\end{equation}}
\newcommand{\bibl}[5]
	{#1, {\it #2} {\bf #3} (#4) #5}

\newcommand{\anti}[1] {${ \ol \mrm #1 }$}
\newcommand{\pair}[1] {${\mrm {#1 \ol #1} }$}
\newcommand{\mrm} {\mathrm}
\newcommand{\srm}[1] {_{\mathrm{#1}}}

\newcommand{\ol} {\overline}
\newcommand{\Ordo}{{\cal O}}

\newcommand{\al} {\bar{\alpha}}

\newcommand{\CF} {{C\srm F}}
\newcommand{\Nf} {{n\srm f}}
\newcommand{\Nc} {N\srm c}

\renewcommand{\d} {{\mrm d}}
\newcommand{\cg} {c_{\mrm g}}
\newcommand{\cq} {c_{\mrm q}}
\newcommand{\cp} {c_{\mrm p}}
\newcommand{\crg} {c_{\mrm r}^{({\mrm g})}}
\newcommand{\crq} {c_{\mrm r}^{({\mrm q})}}
\newcommand{\ctg} {c_{\mrm t}^{({\mrm g})}}
\newcommand{\ctq} {c_{\mrm t}^{({\mrm q})}}
\newcommand{\Yq} {Y_{\mrm q}}
\newcommand{\Yqbar} {Y_{\mrm {\ol q}}}
\newcommand{\gf} {{\cal P}}
\newcommand{\gfg} {{\cal P}\srm{g}}
\newcommand{\gfq} {{\cal P}\srm{q}}
\newcommand{\gfm} {\gf_m}
\newcommand{\Ng} {N\srm g}
\newcommand{\Nq} {N\srm q}
\newcommand{\ngg} {n\srm{gg}}
\newcommand{\Rq} {R^{({\mrm q})}}
\newcommand{\Rg} {R^{({\mrm g})}}
\newcommand{\Rp} {R^{({\mrm p})}}
\newcommand{\antenna}[1]{(\widehat{#1})}

\newcommand{\cAg}{{\cal A}\srm{g}}
\newcommand{\cAq}{{\cal A}\srm{q}}
\newcommand{\cB}{{\cal B}}

\newcommand{\PEfigure}[5]{
\begin{figure}[#1]
\begin{center}
\mbox{\psfig{figure=#2,width=#3\textwidth}}
\caption{\em #4}
\label{f:#5}
\end{center}
\end{figure}
}

\begin{document}

\begin{titlepage}
\begin{flushright}
 NORDITA 00/86 HE\\
 September 2000
\end{flushright}
\vspace{25mm}
\begin{center}
  \Large
  {\bf Multiplicities in Parton and Dipole QCD Cascades} \\
  \vspace{12mm}
  \normalsize
  Patrik Ed\'en\footnote{e-mail eden@nordita.dk}\\
  NORDITA\\
  Blegdamsvej 17, DK-2100 Copenhagen, Denmark\\
\end{center}
\vspace{5cm}
{\bf Abstract:} \\
Evolution equations for multiplicities in QCD cascades can, both in the parton and dipole picture, be used to estimate corrections beyond the formal accuracy of the modified leading log approximation (MLLA).
The differences between the two pictures,  and other uncertainties beyond first order MLLA corrections, are here investigated in some detail. For example, I discuss how some colour suppressed corrections, which cannot be determined without better knowledge of non-perturbative QCD, are related to the question of colour reconnection. A generalized evolution equation for the dipole cascade is also presented.

\end{titlepage}

\section{Introduction}\label{sec:intro}
The modified leading log approximation (MLLA)~\cite{MLLA,pQCDbook} of QCD cascades systematically includes corrections suppressed by a relative factor $\sqrt{\alpha_s}$, where $\alpha_s$ is the coupling of the strong interaction.
Confronting MLLA predictions with data on multiplicities in quark and gluon jets, $\Nq$ and $\Ng$, higher order corrections are found to be important~\cite{OPAL,DELPHI}. These corrections, in particular energy conservation effects, have been estimated by expanding the MLLA evolution equations beyond their order of formal reliability~\cite{Dremin,Ochs,dipolerecoil}. Thus, no exact and systematic inclusion of $\Ordo(\alpha_s)$ effects is available.

One contributor to $\Ordo(\alpha_s)$ corrections is the class of Feynman diagrams where the gluons are not strongly ordered, i.e.\ where two gluons have similar transverse momentum, or where the hardest gluon has a transverse momentum close to the kinematical limit. The iterative cascade formalism is not designed to include such diagrams in a proper way, and thus ``moderately ordered'' gluons introduce uncertainties to the  $\Ordo(\alpha_s)$ corrections, which may have numerical consequences.

The iterative cascade is a branching process, where each emitter can split into two new ones, which continue to emit independently. However, colour suppressed terms, which reflect possible interferences between colour charges due to the finite number of colours, do not fit naturally into this iterative picture. According to~\cite{dipolerecoil}, approximately 10\% of the important energy conservation corrections to $\Ng/\Nq$ are related to these uncertain effects, which cannot be fully determined within the perturbative QCD cascade formalism.  In this paper I discuss their connections to the question of colour reconnection~\cite{WWCR,ZCR}.

A simple way to investigate the uncertainties in the estimates of higher order corrections is to compare two alternative evolution equations, based on the parton cascade~\cite{pQCDbook} and dipole cascade~\cite{GGdipole}, respectively. These are both equivalent within the MLLA accuracy, and their differences are a guide to the uncertainties at higher orders.
In this paper, I compare the theoretical results in the dipole and parton pictures more directly than has been done before, by expanding the dipole evolution equations in $\sqrt{\alpha_s}$, in the same way as has been done for the parton cascade. 

I also present new dipole evolution equations, alternative to the ones in~\cite{GGdipole}. The new derivation is less transparent and the result is mathematically more involved but, in spite of such shortcomings, it is presented for two reasons: First, discussing the differences between the two alternative dipole evolution equations is a powerful way to examine the uncertainties beyond first order MLLA corrections. Second, the recoil corrections, presented in~\cite{dipolerecoil} for average multiplicities only, are here treated in a more general way, suitable for calculations of all multiplicity moments.

The outline of this paper is as follows: In section~\ref{sec:eveqs}, I present the investigated evolution equations. In section~\ref{sec:comp}, their predictions on some multiplicity quantities are compared. The differences and other uncertainties at higher order corrections are discussed in section~\ref{sec:terms}. In section~\ref{sec:CR} a relation to colour reconnection is discussed. The paper ends with a summary in section~\ref{sec:concl}.

\section{Evolution Equations}\label{sec:eveqs}
In this section, I list three alternative evolution equations for the multiplicity distribution in QCD cascades. Though this paper focuses on average multiplicities, I will when possible give the evolution equations for the more general generating functions.
The generating function to a distribution, $P(\lambda)$, of some quantity $\lambda$, is defined as
\eqbe \gf(z) \equiv \int\d \lambda P(\lambda)(1+z)^\lambda. \eqen
While the distribution for two indepently contributing sources is described by a convolution, the corresponding generating function is simply multiplicative, 
\eqbe P_{a+b}(\lambda) = \int\d\lambda_1\d\lambda_2P_a(\lambda_1)P_b(\lambda_2)\delta(\lambda-\lambda_1-\lambda_2) \Rightarrow \gf_{a+b}(z) =\gf_a(z)\gf_b(z). \eqen
(In case of a discrete variable, e.g.\ multiplicity, the integrals are replaced by sums, and the $\delta$ distribution by a Kronecker $\delta$.)
Expanding $\gf$ in powers of $z$ gives different moments of the multiplicity distribution, and thus knowledge of $\gf$ gives full knowledge of the distributions.

\subsection{Parton Equation}\label{sec:parteqs}
The evolution equations for generating functions for the multiplicity distributions are in the parton cascade picture~\cite{pQCDbook} given by
\eaABCbe{peveq}
\ealine{\gf_G'(l)}{=}{\int\d xK_G^G(x)2\al[\gf_G(l+\ln(x))\gf_G(l+\ln(1-x))-\gf_G(l)]}\\
\ealine{}{}{+\frac{\Nf}{2\Nc}\int\d xK_G^F(x)2\al[\gf_F(l+\ln(x))\gf_F(l+\ln(1-x))-\gf_G(l)],}\\[6mm]
\ealine{\gf_F'(l)}{=}{\frac{\CF}{\Nc}\int\d xK_F^G(x)2\al[\gf_G(l+\ln(x))\gf_F(l+\ln(1-x))-\gf_F(l)].}
\eaABen{-20mm}{8mm}
Here $\gf_F$ represents a quark jet and $\gf_G$ a gluon jet. $x$ is the momentum fraction of the emitted gluon (or quark, in case of g$\to$\pair q splitting), $\Nf$ is the number of active quark flavours, $\Nc=3$ is the number of colours and $\CF=(\Nc/2)(1-\Nc^{-2})=4/3$. The dependence on the auxiliary variable $z$ is suppressed in the notation, and as kinematical variable is used $l = \ln(Q/\Lambda)$, where $Q$ is the virtuality of the jet and $\Lambda$ is the scale parameter of the perturbative cascade.
The running coupling, depending on the transverse momentum $k_\perp$, is written as
\eqbe \al = \frac{\Nc\alpha_s}{2\pi} = \frac{2\Nc}{\beta_0\kappa}(1-\frac{2\beta_1\ln(\kappa)}{\beta_0^2\kappa})+\Ordo(\kappa^{-3}), \label{e:aldef} \eqen
where
\eqbe \kappa=\ln(k_\perp^2/\Lambda^2), \label{e:kdef} \eqen
\eqbe \beta_0\equiv\frac13\left(11\Nc-2\Nf\right),~~~~\beta_1\equiv \frac13\left(17\Nc^2-\Nf(5\Nc+3\CF)\right). \eqen
The splitting kernels $K$ can be written
\eqbe K_F^G(x) =\frac{1+(1-x)^2}{x},~~~K_G^G(x)=\frac{1+(1-x)^3}{x},~~~K_G^F(x)=x^2+(1-x)^2. \label{e:Kdefs}\eqen
$K_G^G$ must be chosen so that the sum
\eqbe \frac12[K_G^G(x)+K_G^G(1-x)]=\frac{1-x}x +\frac{x}{1-x}+x(1-x) \label{e:APsum} \eqen
agrees with the Altarelli-Parisi splitting function for g$\to$gg. 
The choice made here is convenient for comparison with the dipole formalism.

\subsection{Dipole Equation}\label{sec:dipeqs}
In~\cite{dipolerecoil}, the dipole evolution equations presented in~\cite{GGdipole} are supplemented with recoil corrections. Only the average effect of recoils is considered, and it is therefore preferable to use the equations for the average multiplicity only, and not higher moments. The result is
\eaABCbe{deveq}
\ealine{\Ng''(L)}{=}{\al(L-\cg)\left[1-\crg\al(L-\cg)\right]\Ng(L-\cg)}\\[2mm]
\ealine{}{}{+\frac{\Nf}{3\Nc}\frac{\d}{\d L}\{\al(L)[2\Nq(L-\cp)-\Ng(L)]\},}\\[6mm]
\ealine{\Nq''(L)}{=}{\frac{\CF}{\Nc}\al(L-\cq)\left[1-\crq\al(L-\cq)\right]\Ng(L-\cq).}
\eaABen{-20mm}{7.5mm}
Here $\Ng(\Nq)$ is the mean multiplicity of a gluon(quark) jet, defined as one hemisphere of a gg (\pair q) system stemming from a point source, with invariant mass $\sqrt s$. This scale is represented by the logarithmic variable
\eqbe L=\ln(s/\Lambda^2). \eqen 
The number $\cg$($\cq$) represents corrections related to the non-singular part of the gluon emission kernel $K_G^G$($K_F^G$), in the following referred to as ``kernel corrections''. The number $\crg(\crq)$ is related to recoil effects in a gg(\pair q) dipole, while $\cp$ reflects energy conservation in pair production g$\to$\pair q. The values are
\eqbe \cg = \frac{11}6,~~~~\cq=\frac32,~~~~\cp=\frac{13}6, \label{e:cdefs} \eqen
\eqbe \crg= 2(\frac{\pi^2}6-\frac{49}{72}),~~~\crq= -\frac{2}{\Nc^2}(\frac{\pi^2}6-\frac{5}{8}). \label{e:crdefs} \eqen

\subsection{Generalized Dipole Equation}\label{sec:gendipeqs}
In appendix, I derive dipole evolution equations alternative to Eqs~(\ref{e:deveq}). The recoil corrections are here treated in  way suitable for calculations of all multiplicity moments. I therefore refer to these formulas as the generalized dipole evolution equations.
The equations are
\eaABCbe{neveq}
\ealine{[\ln(\gfg(L+\cg))]'}{=}{\cB(L)+ \al(L)\gfg(L+\cg)\exp(-\cg\cB(L))\sum_{n=1}\Rg_n\cB^n(L)+}\\[2mm]
\ealine{}{+}{\!\! \frac{\Nf\al(L)}{3\Nc}\!\!\left[\!\frac{\gfq(L\!+\!\cq)}{\gfg(L\!+\!\cg)}\!\exp[(\cg\!-\!\frac{2\CF}{\Nc}\cq)\cB(L)]\!\sum_{n=0}\Rp_n{\cB}^n(L)\! -\! 1 \right],}\\[6mm]
\ealine{[\ln(\gfq(L+\cq))]'}{=}{\!\!\frac{2\CF}{\Nc}\!\left[ \cB(L) + \al(L)\gfg(L+\cg)\exp(-\cg \cB(L))\sum_{n=1}\Rq_n \cB^n(L)\right],~~~~}\\[6mm]
\ealine{\cB'(L)}{=}{\al(L)[\gfg(L+\cg)\exp(-\cg\cB(L))-1].}
\eaABCen{-31mm}{10mm}{5.5mm}
Here $\cB$ is an auxiliary generating function, whose moments in the generating function variable $z$ can be eliminated, replaced by moments of the physically more relevant functions $\gfg$ and $\gfq$, which represent the multiplicity in a gg and \pair q dipole, respectively. Note that a gg system consists of two gg dipoles, one from each triplet charge to the matching anti-triplet charge of the other gluon. The numbers $R^{(i)}_n$, defined in Eqs~(\ref{e:Rgpdef}),~(\ref{e:Rqpdef}) and~(\ref{e:Rppdef}), reflect energy conservation. In particular, we have $c^{(i)}_{\mrm r}=-R_1^{(i)}$.

\section{Comparison}\label{sec:comp}
Both the parton and dipole pictures have been used to predict multiplicity observables. In~\cite{Dremin} the generating functions of Eqs~(\ref{e:peveq}) are expanded $\gf(l+\varepsilon) = \gf(l)+\varepsilon\gf'(l)+...$, to get parton cascade evolution equations expanded  in $\sqrt{\al}$. 
In~\cite{dipolerecoil}, the dipole evolution equations, Eqs~(\ref{e:deveq}), are combined to
\eqbe \Ng'(L+\cg-\cq) = \frac{\Nc}{\CF}[1-(\crg-\crq)\al(L)]\Nq'(L). \label{e:recoilcombined}   \eqen
Using the experimentally well determined $\Nq$ as input, and a boundary condition scale $L_0$ where $\Nq(L_0)=\Ng(L_0)$, this equation is used to predict $\Ng$ for scales $L>L_0$. These parton and dipole approaches are confronted with data in~\cite{DELPHI}.

In this section,  all evolution equations are expanded in $\sqrt{\al}$ in the same way, thus enabling
a more direct comparison of the cascade pictures. To explore the differences, it is enough to consider average multiplicities, with corrections up to second order in $\sqrt{\al}$.

With numbers $a_i$ and $r_i$ defined as in~\cite{Dremin} from
\eabe
\frac{\Ng'(L)}{\Ng(L)} & = & \sqrt{\al}\left(1-2a_1\sqrt{\al}-4a_2\al+\Ordo(\al^{3/2})\right),\\
\frac{\Ng(L)}{\Nq(L)} & = & r_0\left(1-2r_1\sqrt{\al}-4r_2\al+\Ordo(\al^{3/2})\right),
\eaen
the evolution equations for average multiplicities can be written in the following form, covering corrections up to second order,
\eaABCbe{Nbis}
\ealine{\Ng''}{=}{\al\Ng-(\cg+\frac{\Nf}{3\Nc^3})(\al\Ng)' +4\left[\frac{\Nf(r_1-\frac{\cp}2)}{r_03\Nc}+g_2\right] \al^2\Ng,}\\[2mm]
\ealine{\frac{\Nc}{\CF}\Nq''}{=}{\al\Ng-\cq(\al\Ng)'+4q_2 \al^2\Ng.}
\eaABen{-17mm}{3mm}
\begin{table}[tb]
\begin{center}
\begin{tabular}{|c|c|c|c|}
\hline
 & parton & dipole & generalized dipole \\
\hline
&&&\\[-2mm]
$\displaystyle 4g_2$ & $4v_2$ & $\frac12 \displaystyle \cg^2-\crg$ & $\displaystyle -\crg-\cp\frac{\Nf}{2\Nc^3}$\\[2mm]
$\displaystyle 4q_2$ & $\displaystyle -4v_8-4\frac{v_{10}}{r_0}$ & $\frac12 \displaystyle \cq^2-\crq$ & $\frac12 \displaystyle (\cq^2-\cg^2)-\crq-\cg\frac{\Nf}{3\Nc^3}$\\[3mm]
\hline
\end{tabular}
\caption{\em Table with second order coefficients which differ between the considered cascades. The values of the quantities are given in the text. Their interpretation is discussed in section~\protect{\ref{sec:terms}}. Second order terms common to all cascades are given in Eq~(\ref{e:Nbis}).}\label{t:coeffs}
\end{center}
\end{table}
The different cascade expressions for $g_2$ and $q_2$ are shown in Table~\ref{t:coeffs}.
The coefficients $v$, defined in~\cite{Dremin}, reflect recoil effects in the parton cascade. The values are
\eqbe v_2 = \frac{67}{36}-\frac{\pi^2}6,~~~v_8=-\frac78,~~~v_{10}=\frac{\pi^2}6-\frac58. \eqen
Matching orders of $\sqrt{\al}$ in Eqs~(\ref{e:Nbis}), we get
\eabe r_0 = \frac{\Nc}{\CF},&&2a_1 = \frac{\cg}2+\frac{\Nf}{6\Nc^3}-\frac{\beta_0}{8\Nc},~~~~r_1 = \frac{\cg-\cq}2+\frac{\Nf}{6\Nc^3},\\
2a_2 & = & -(a_1+\frac{\beta_0}{8\Nc})(a_1+\frac{\beta_0}{4\Nc})-\frac{\Nf}{r_03\Nc}(r_1-\frac{\cp}2)-g_2,\\
r_2 & = & r_1(\frac{\cq}2-a_1)-\frac{\Nf}{r_03\Nc}(r_1-\frac{\cp}2)-g_2+q_2.
\eaen
\begin{table}[tb]
\newcommand{\m}{\hphantom{$-$}}
\begin{center}
\begin{tabular}{|c|ccc|ccc|}
\hline
& \multicolumn{3}{c|}{$a_2$} & \multicolumn{3}{c|}{$r_2$} \\
$\Nf$ & parton & dipole & gen.\ dip.\ &parton & dipole & gen.\ dip.\ \\
\hline
&&&&&&\\[-2mm]
$3$ &$ -0.379 $ &$-0.240$&$-0.015$&$ 0.426 $ &$0.620$&$0.633$\\
$4$ &$ -0.339 $ &$-0.200$&\m$0.030$&$ 0.468 $ &$0.663$&$0.680$\\
$5$ &$ -0.300 $ &$-0.161$&\m$0.074$&$ 0.510 $ &$0.705$&$0.727$\\
\hline
\end{tabular}
\caption{\em Numerical differences on second order corrections to the anomalous dimension and the multiplicity ratio $\Ng/\Nq$.}\label{t:numbers}
\end{center}
\end{table}
\PEfigure{tb}{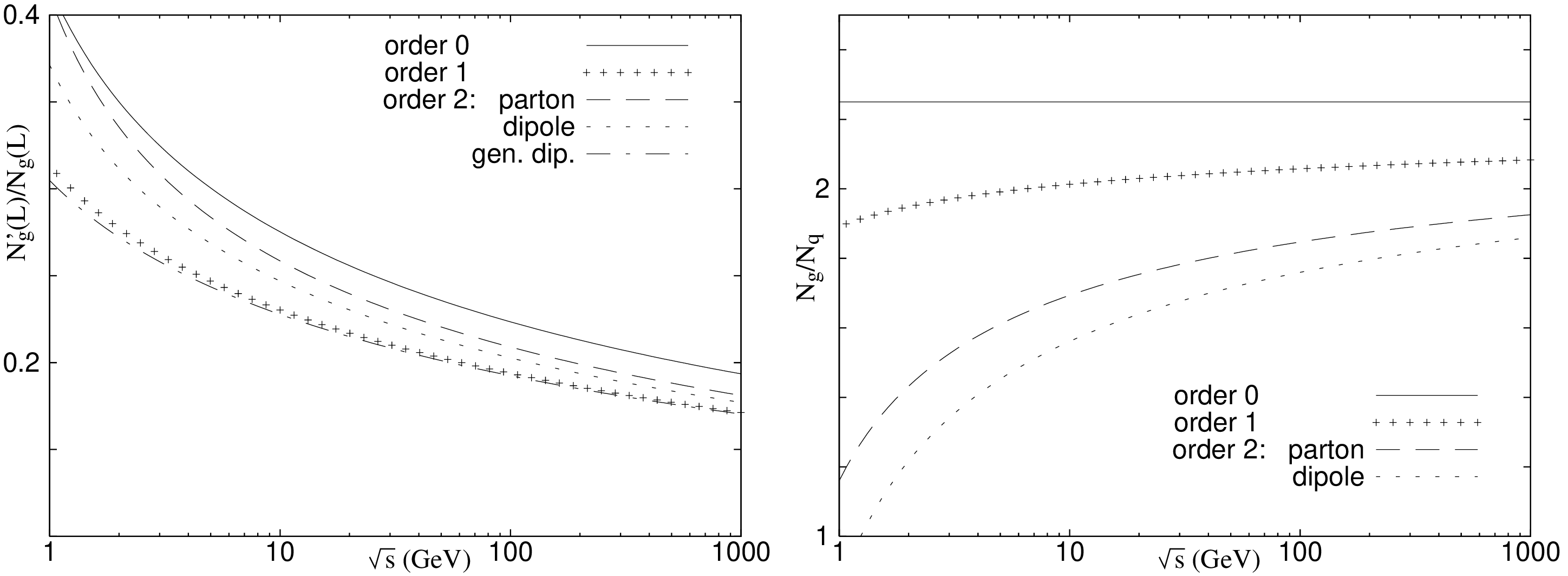}{0.98}{Multiplicity ratios to different orders in $\sqrt{\al}$. The chosen example is $\Nf=4$ and $\Lambda=0.22$GeV. $\al$ is given by Eq~(\ref{e:aldef}). {\bf left:} Anomalous dimension $\Ng'/\Ng$ as $\sqrt{\al}$ (solid line), $\sqrt{\al}(1-2a_1\sqrt{\al})$ (crossed) and $\sqrt{\al}(1-2a_1\sqrt{\al}-4a_2\al)$ for the parton equation (dashed), dipole equation (dotted) and generalized dipole equation (dash-dotted). {\bf right:} $\Ng/\Nq$  as $r_0$ (solid line), $r_0(1-2r_1\sqrt{\al})$ (crossed), and $r_0(1-2r_1\sqrt{\al}-4r_2\al)$ for the parton picture (dashed) and dipole picture (dotted). }
{results}

The numerical impact of the equation differences to the anomalous dimension $\Ng'/\Ng$ and multiplicity ratio $\Ng/\Nq$ is presented in Table~\ref{t:numbers} and Fig~\ref{f:results}. We conclude that there is a considerable difference between the parton and dipole pictures, also when precisely the same procedure has been invoked to reach predictions on physical observables.
The uncertainties represented by differences in $a_2$ and $r_2$ are larger than the third order corrections calculated for the parton formalism~\cite{Dremin}.

\section{Discussion of Corrections up to Second Order}\label{sec:terms}
\PEfigure{tb}{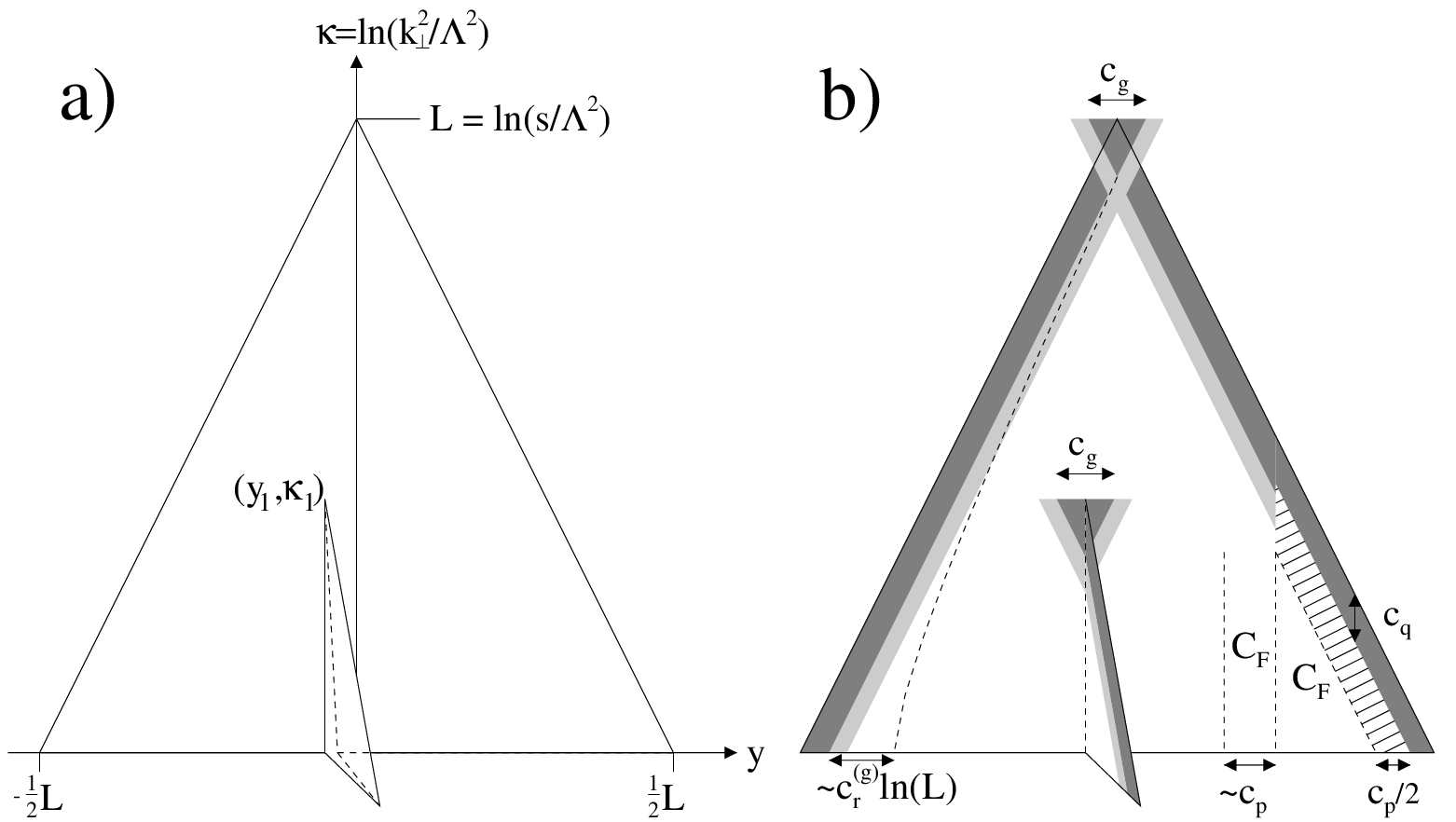}{0.98}{{\bf a)} The phase space for gluon is, to leading order, a triangluar region in the $y,\kappa$ plane. A gluon emitted at $y_1,\kappa_1$ enlarges the phase space with a double-sided fold of height $\kappa_1$. This implies that the triangular phase space represents the leading order relation $\Ng''=\al\Ng$. {\bf b)} Illustration of some subleading corrections: The dark strips illustrate effective phase space reductions representing kernel corrections. The light-gray strips represent the phase space for gluon splittings. If ${\mrm{g}}\to$\pair q, the quark colour factor $\CF$ applies to part of the dipole phase space. The phase space reduction due to the lower energy of the produced quark is illustrated with the dashed strip.}{dipoleterms}
In this section, I discuss the different terms in Eqs~(\ref{e:Nbis}) and Table~\ref{t:coeffs}, and also some not included anywhere yet.
Many corrections can be visualized with the logarithmic phase space triangle.
Following the colour flow connecting the partons, we label the original ones 1 and 3, and the emitted gluon 2, and let $x_i$ be the scaled energy $2E_i/\sqrt s$ of parton $i$.
 With kinematical variables
\eqbe y=\frac12\ln\left(\frac{1-x_1}{1-x_3}\right),~~~~k_\perp^2=s(1-x_1)(1-x_3), \label{e:kappadef}\eqen
the constraint $2k_\perp\cosh (y)<\sqrt s$ implies
\eqbe |y| \le y_{\max} \equiv \ln(\frac{\sqrt{s}}{k_\perp})+\ln(\frac{1+\sqrt{1-4k_\perp^2/s}}{2})  \approx \frac12(L-\kappa). \label{e:ymdef} \eqen
Thus the  allowed phase space for gluon emission is approximately a triangular region in the $y,\kappa$-plane, cf.\ Fig~\ref{f:dipoleterms}a.
As an introduction to this phase space picture, I discuss how it illustrates leading order results and first order corrections, thereafter continuing with second order corrections of increasing uncertainty. In the following, the figure represents a gg dipole.

\subsection{Simpler Terms}\label{sec:simpleterms}
After the emission of a gluon at $y_1,\kappa_1$, the additional phase space for further emissions can be represented by a fold of height $\kappa_1$, as in Fig~\ref{f:dipoleterms}a. The fold contributes with a multiplicity $\Ng(\kappa_1)$. Neglecting kernel corrections and gluon splittings, this implies
\eqbe \Ng(L)=\int^L\d\kappa\int_{\frac12(\kappa-L)}^{\frac12(L-\kappa)}\d y \al(\kappa)\Ng(\kappa), \label{e:evintegral} \eqen
giving the leading order result  $\Ng''=\al\Ng$  in Eq~(\ref{e:Nbis}a). Thus, the triangular phase space in Fig~\ref{f:dipoleterms}a illustrates the leading order term.

In Fig~\ref{f:dipoleterms}b, some possible corrections up to second order are illustrated. Strips of finite width correspond to integrals like Eq~(\ref{e:evintegral}), but with the $y$ interval constant, resulting in terms in Eq~(\ref{e:Nbis}) proportional to $(\al\Ng)'$. A finite area, where also the $\kappa$ limits are independent of $L$, represent terms proportional to $(\al\Ng)''\to \al^2\Ng$.

At the left edge, the dark strip of width $\cg/2$ illustrates an effective phase space reduction due to kernel corrections, and the light-gray strip represents the phase-space for gluon splittings. To avoid drawing two strips on top of each other, the ``splitting strip'' is here drawn just inside the ``$\cg$ strip''. Together, they illustrate the first order correction $-(\cg+\Nf/3\Nc^3)(\al\Ng)'$ in Eq~(\ref{e:Nbis}a). In a \pair q dipole, a corresponding strip represents  $-\cq(\al\Ng)'$ in Eq~(\ref{e:Nbis}b).

At the right edge, the effects of gluon splittings is illustrated. Below the transverse scale of the gluon splitting, the colour factor is reduced from $\Nc/2$ to $\CF$. The strips at the edge are changed, and due to the reduced energy of the quark, there is also a strip of width $\cp/2$ removed. These changes of strips illustrate first order corrections to the gluon splitting, which in itself is a first order correction, thus representing the second order term $(r_1-\cp/2)(4\Nf/r_03\Nc)\al^2\Ng$ in Eq~(\ref{e:Nbis}a).

The colour factor reduction need not cover precisely the triangular fold below the gluon splitting. The border for colour suppression may be moved a finite distance, whose mean could be related to the average energies of the \pair q pair, and thus to $\cp$. In the generalized dipole equation, this mean is $3\cp/4$, giving the second order correction $-(\cp\Nf/2\Nc^3)\al^2\Ng$ in Table~\ref{t:coeffs}.

As discussed in the introduction, the top of the triangle and the surroundings of a gluon fold top are regions where the validity of the iterative cascade formalism is uncertain. However, second order corrections related to ``moderately ordered gluons'', emitted in these regions, are of some numerical importance. This is e.g.\ observed in~\cite{dipoletop}, discussing the triangle top correction.

The $\cg$ strips of height $L$ overshoot at the triangle top. Re-inserting a finite area, we get a second order correction $(\cg^2/2)\al^2\Ng$. Assuming the  $\cg$ strips at the edge of gluon folds to extend into the background triangle, we get second order corrections which precisely cancel the ones at the top of the gg dipole, while the top of a \pair q dipole is different.
From the coefficient $q_2$ in Table~\ref{t:coeffs}, we find that the dipole equation includes the $(c^2/2)$-terms at the top, but not at a gluon fold, while the generalized dipole equation includes both. This is the main difference between the two equations, and we noted in section~\ref{sec:comp} its numerical effect on the anomalous dimension $\Ng'/\Ng$.

Gluon splittings may also be ``moderately ordered'' and it is possible to have corrections represented by  a V-shaped splitting area overshooting at the top of a gg dipole, and a compensating V-shaped area around the top of a gluon fold. Looking for $\Nf$ dependence in $q_2$, we see that these gluon splitting assumptions differ between the two dipole equations, but that the effects are numerically small.

Note that the discussion here concerns {\em possible} second order corrections. One cannot conclude that the equation including the most should be more appropriate. 
The parton cascade treatment of moderately ordered gluons is discussed at the end of the next subsection.

\subsection{Very Hard Gluons}\label{sec:top}
There are possible corrections related to emissions of high-$k_\perp$ gluons, not included in any of the evolution equations. The corrections are not numerically insignificant, but as the iterative cascade formalism is very uncertain for these extreme gluons, I leave the results in section~\ref{sec:comp} unchanged. The discussion here is mainly meant to further illustrate uncertainties beyond first order MLLA corrections.

\subsubsection*{In the Dipole Cascade}
In order to take a closer look at high-$k_\perp$ corrections, we note that the emission densities off \pair q and gg dipoles are
\eqbe \d n\srm{q\ol q}  =  \frac{2\CF}{\Nc}\al\d\kappa\d y\frac{x_1^2+x_3^2}2,~~~~\d \ngg  =  \al\d\kappa\d y\frac{x_1^3+x_3^3}2. \label{e:dn} \eqen
In the limit $k_\perp^2\ll s$, the factors $\d y(x_1^n+x_3^n)/2$ equal $K_F^G(x)\d x$ and $K_G^G(x)\d x$, respectively. 

In the dipole evolution equations, the kernel corrections are represented by  constant rapidity range reductions $\cg$ and $\cq$. For very high-$k_\perp$ gluons, this is only approximately correct.
The effective phase space differences, whose main contributions are at high $\kappa$, have magnitude (index t stands for ``top correction'')
\eabe \ctq  & \equiv & \lim_{L\to\infty}\left\{\int_0^{L-\ln 4}\d \kappa\int_{-y_{\max}}^{y_{\max}}\d y\frac{x_1^2+x_3^2}2 - \frac12(L-\cq)^2\right\}\nonumber\\
& = & \cq(2-\frac{\cq}2)-\ln^2(2)-\sum_{n=1}\frac{2}{2^nn^2}\approx 0.23,\\
\ctg & = & \frac1{18}+\cg(2-\frac{\cg}2)-\ln^2(2)-\sum_{n=1}\frac2{2^nn^2}  \approx 0.40. \label{e:ctgdef} \eaen
These top corrections modifies $a_2$ and $r_2$ by
\eqbe \Delta a_2 = -\frac{\ctg}8 \approx -0.050,~~~\Delta r_2 = \frac{\ctq-\ctg}4 = -\frac1{24}\approx -0.042.\eqen
Comparing with e.g.\ the differences in $a_2$ between cascade formalisms, shown in Table~\ref{t:numbers}, we note that these top corrections, if trustworthy, are not numerically negligeble.

If the only constraint on the gg dipole emission density is to reproduce the Altarelli-Parisi splitting function in its region of validity, the gg kernel in Eq~(\ref{e:dn}) is not unique. We may generalize $x_1^3+x_3^3$ to
\eqbe x_1^3+x_3^3+x_2[f(x_2)-f(1-x_2)]+\frac{k_\perp^2}{s}g(\frac{k_\perp^2}{s},x_2). \label{e:ggkernelgeneral} \eqen
This kernel is symmetric in indices 1 and 3. The factor $(k_\perp^2/s)g$  does not affect $K_G^G$, provided $g$ is regular. The generalized kernel in Eq~(\ref{e:ggkernelgeneral}) just adds a term $f(x)-f(1-x)$ to $K_G^G(x)$, which cancels against the corresponding term in $K_G^G(1-x)$, leaving the sum in Eq~(\ref{e:APsum}) unaffected. (For consistency, the kernel for a mixed (qg) dipole must be modified from $x\srm q^2+x\srm g^3$ to $x\srm q^2+x\srm g^3+(1-x\srm g)[f(x_2)-f(1-x_2)]+\frac{k_\perp^2}{2s}g(\frac{k_\perp^2}{s},x_2)$. The additional terms vanish when $x\srm g\to 1$, leaving $K_F^G$ unchanged.)

To examine the numerical impact of this possible kernel generelization, consider the simple case $g=0$, $f(x_2)=\chi x_2$, where $\chi$ is a free paramter.
This modification adds $\chi$ to the top correction $\ctg$ and $\chi/2$ to the recoil coefficient $\crg$. Constraining the kernel in Eq~(\ref{e:ggkernelgeneral}) to be positive everywhere in phase space, not to overthrow the ansatz of independently emitting dipoles, implies $-2 <8\chi <(35+13\sqrt{13})$.
With minimum $\chi $, the change in $a_2$ is moderated from $-0.050$ to $-0.034$, and the change in $r_2$ from $-0.042$ to $-0.010$, but there is no a priori argument for this minimum value to be appropriate. Instead letting $\chi$ approach its maximum value $\sim 10$, would have drastic effects on the second order result.
However, the possible choices in Eq~(\ref{e:ggkernelgeneral}) are constrained by data on e.g.\ multiplicity distributions, which are very well described by the cascade formalism, and four-jet characteristics in \pair q$\to$\pair qgg, which are best described by the QCD matrix elements, but reasonably well reproduced by cascades~\cite{MEmatch}. Reversing the discussion, the degrees of freedom in Eq~(\ref{e:ggkernelgeneral}) could be used to search for an improved simultaneous fit to multiplicity and 4-jet properties.

\subsubsection*{In the Parton Cascade}
In the parton picture, a two-parton system is described as two independent jets, with opening angle $\pi/2$. However, in~\cite{Ochs} it was pointed out that this leads to an approximate treatment of very hard gluon emission, which has numerical consequences.
To improve parton cascade calculations they use, for the first gluon off a \pair q system, the \pair q emission density in Eq~(\ref{e:dn}).

A similar well defined matrix element is not available for gg $\to$ ggg. If trying to include similar corrections in the gg case, by letting the hardest gluon be emitted in a dipole-like manner, rather than from two independent gluon jets, the parton formalism would inherit the gg dipole high-$k_\perp$ uncertainties discussed above.

\subsection{Recoil Terms}\label{sec:recoils}
In the dipole picture, a gg dipole takes recoils from  emissions in the neighbouring dipoles, and thus the phase space is reduced. On average, the rapidity reduction at transverse momentum scale $\kappa$ is $\crg\int_\kappa^{L-\cg}\d l\al(l)$ (dashed line at left edge of Fig~\ref{f:dipoleterms}b). This gives the second order recoil correction $-\crg\al^2\Ng$ in Table~\ref{t:coeffs}. 

\PEfigure{tb}{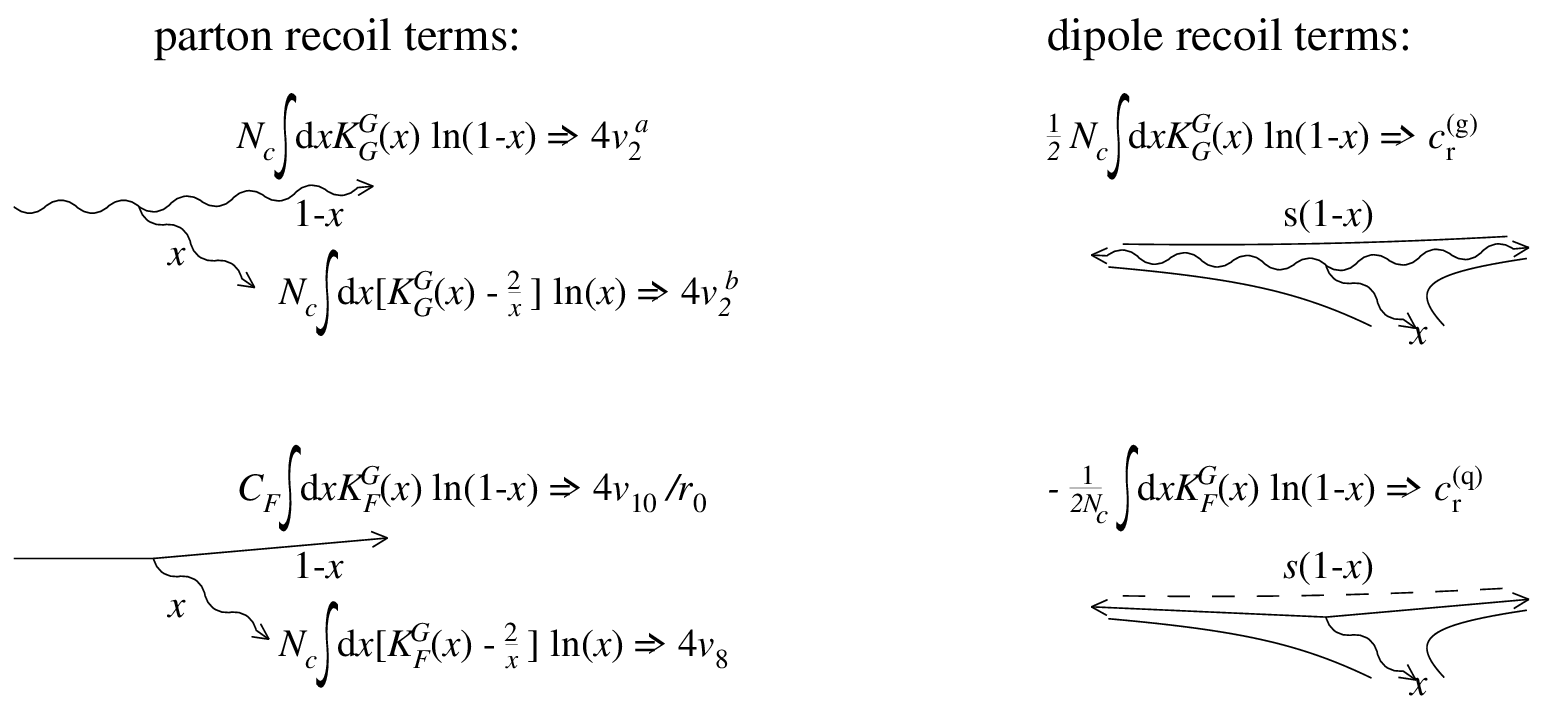}{0.9}{Comparing recoil terms. {\bf left:}  In the parton cascade, recoils on the emitting parton lead to the recoil terms $4v_2^a$ and $4v_{10}/r_0$. The energy fraction $x$ of the emitted gluon, combined with  kernel corrections, leads to the recoil terms $v_2^b$ and $v_8$. {\bf right:} In the dipole picture of ${\mrm{gg}}\to{\mrm{ggg}}$, the squared mass of the ``spectator diple'' is reduced by a factor $1-x$. There is no corresponding spectator dipole in case of emission from a \pair q pair. However, a colour suppressed term can be represented by a dipole between the q and \anti q, whose contribution is weighted with the negative factor $-1/\Nc^2$. This is illustrated by the dashed line in the lower right figure. 
}{branchfigs}
A more complete set of recoil terms is illustrated in Fig~\ref{f:branchfigs}.
There,  $v_2$ is split in two parts,
\eqbe v_2^a=\frac{\pi^2}6-\frac{49}{72},~~~v_2^b=-\frac{85}{72} \label{e:v2split}, \eqen
which have a direct correspondance to the quark terms $v_{10}/r_0$ and $v_8$, respectively.
$v_2^b$ and $v_8$ come from a combination of kernel corrections with energy conservation on the emitted gluon, which has no correspondance in the dipole picture.

Recoils in the parton picture, acting on an emitting gluon, corresponds in the dipole picture to recoils on a spectator gg dipole. Since the colour factor is $\Nc$ for gluons and $\Nc/2$ for gg dipoles, this leads to the factor two difference between $4v_2^a$ and $\crg$.

The recoil factor difference $4v_{10}/r_0-\crq$ is related to the same colour factor $\Nc/2$ as is $4v_2^a-\crg$. 
In the emission density off a \pair qg system, a colour suppressed term can be represented by a dipole between the q and \anti q, whose contribution is weighted with the negative factor $-1/\Nc^2$. Recoils in the dipole picture act on this colour correction dipole. However, since it can not be regarded an independent emitter, full understanding of its effects requires better knowledge of non-perturbative QCD, e.g.\ questions related to so called colour reconnection. This is further discussed in section~\ref{sec:CR}.

In the dipole model, specific assumptions about the effects of the negative dipole are made. Using the antenna symbol $\antenna{ij}$ to represent a dipole between partons $i$ and $j$~\cite{pQCDbook}, the emission density from a \pair qg state can be written
\eqbe \d n \propto \al[\antenna{{\mrm{qg}}}+\antenna{{\mrm{g\ol{q}}}}-\frac 1{\Nc^2}\antenna{{\mrm{q\ol{q}}}}]. \eqen
The antenna $\antenna{ij}$ has collinear divergences in the directions of both its partons. To arrive at a parton cascade picture, each antenna is split in two parts, $\antenna{ij}=P_{ij}+P_{ji}$, where $P_{ij}(P_{ji})$ has collinear divergences in the direction of parton $i(j)$ only, and the emission density is rewritten in the form
\eqbe \d n \propto \frac{\alpha_s}\pi[\CF (P\srm{qg}+P\srm{\ol{q}g})+\Nc P\srm{g\ol{q}}] +\al\{P\srm{gq}-P\srm{g\ol{q}}\} -\frac{\al}{\Nc^2}\{\antenna{\mrm{q\ol{q}}}-P\srm{qg}-P\srm{\ol{q}g}\} . \eqen
The collinear divergences cancel inside the curly brackets, and within first order MLLA accuracy, it is appropriate to neglect them. We note, however, that possible higher order corrections from the second curly bracket, depending on the negatively weighted $\antenna{\mrm{q\ol{q}}}$ antenna, cannot be fully determined without better knowledge of non-perturbative QCD.

\section{Colour Reconnection}\label{sec:CR}
In the perturbative cascade, the dipoles, or the angular ordering constraints on partons,  associates
each triplet charge with a particular anti-triplet charge. Due to the finite number of colours, this colour topology is not unique. Interference between different topologies are possible, though suppressed.
Furthermore, in the two most successful phenomenological hadronization models, a massless relativistic string~\cite{string} or a chain of clusters~\cite{cluster} connects triplet and anti-triplet charges. It is not necessary that this non-perturbative colour topology, which in the models has strong influence on the hadronic final state, agrees with the perturbative colour topology.

The question of colour reconnection, where the perturbative or non-perturbative colour topology changes from the one defined by planar Feynman diagrams,  has acquired quite some attention because of its possible effects on W mass measurements in $e^+e^-\to {\mrm{W}}^+{\mrm{W}}^-$~\cite{WWCR}.
To gain more insight to the problem,
an observable in Z decays has been proposed~\cite{ZCR}. In  events with a very hard gluon emission, due to which the quark and anti-quark are essentially collinear, colour reconnection could imply that the hard gluon forms a colour singlet system with some softer gluons nearby in phase space. If so, there need not be any colour flow connecting the \pair q pair and the hard gluon. This implies a multiplicity depletion at central rapidities along the thrust axis.

In data from Z decays~\cite{OPALCR}, no effect of colour reconnection is seen. Default Monte Carlos are in excellent agreement with data, while a colour reconnection model~\cite{LeifCR} is disfavoured.
However, in the considered events, a high-$k_\perp$ gluon is emitted, and for subsequent emissions, $s_{\mrm{q\ol q}}$ is small. According to the recoil picture where $s\srm{q\ol q}$ determines the phase space for colour factor reduction, small $s\srm{q\ol q}$ implies that the q and \anti q emit coherently as an octet charge in most of the available phase space. In the perturbative phase, this implies that only a second  gluon essentially collinear with the q or \anti q could reconnect to form a singlet with the hard gluon in the opposite hemisphere. After such a reconnection, there is still colour flow over central rapidities and there is no multiplicity depletion expected. This possible \pair q coherence effect is not considered in the colour reconnection models confronted with data~\cite{ZCR,LeifCR}. Thus, the uncertainties about recoils on the negatively weighted dipole might dilute the discriminating power of the suggested observable.

\section{Summary}\label{sec:concl}
Results from parton and dipole evolution equations, expanded one order beyond the formal accuracy of the modified leading log approximation, differ noticeably. This illustrates the theoretical uncertainty at this order. The difference between cascade pictures is larger than the third order correction term calculated in the parton cascade~\cite{Dremin}. Data on $\Ng/\Nq$~\cite{OPAL,DELPHI} seem to favour the smaller result in the dipole cascade.

One important source of uncertainties are ``moderately ordered gluons'', i.e.\ two gluons with similar $k_\perp$, or a first hard gluon with $k_\perp$ close to the kinematical limit. In particular, the emission density for very hard gluons off gluonic systems is not fully determined by the constraint on soft gluon emission given by the Altarelli-Parisi splitting function. This gives an opportunity to search for a emission kernel which improves the simultaneous fit to multiplicity properties (very well described by cascades) and four-jet characteristics (better described by matrix elements).

Part of the recoil effects are truly non-perturbative, related to negative colour suppressed terms which can not be represented by independently emitting sources. There are possible relations between these recoil effects and the question of colour reconnection.

In this paper a generalized dipole evolution equation is presented, which includes energy conservation effects in a more general way than has been done before in the dipole formalism~\cite{dipolerecoil}. This equation is suitable to calculate higher order corrections to all multiplicity moments, not only average multiplicity, in the dipole cascade picture.

\subsection*{Acknowledgments}
I thank prof.\ G\"osta Gustafson for fruitful discussions.

\appendix
\section{The Generalized Dipole Evolution Equation}\label{sec:newderivation}
In this appendix, I derive a generalized dipole evolution equation alternative to the one presented in ref~\cite{GGdipole}. 
Since the dipole cascade is ordered in $k_\perp$,
I introduce a generating function $\gf(L,\kappa)$, where $L$ and $\kappa$ are logarithms of the dipole energy and maximum allowed $k_\perp$, respectively.

\subsubsection*{Gluon Dynamics}
In gluon dynamics, based on $\d\ngg$ in Eq~(\ref{e:dn}), the generating function for dipoles satisfies
\eabe
\gfg^2(L,\kappa) &=& \int_{\kappa_c}^\kappa\d \kappa'2\al(\kappa')S(L,\kappa,\kappa')\int_{-y_{\max}}^{y_{\max}}\d y\frac{x_1^3+x_3^3}{2}\gfg(L_{12},\kappa')\gfg(L_{23},\kappa')\gfg(L_{13},\kappa') \nonumber\\&&+ S(L,\kappa,\kappa_c)\gfg^2(L,\kappa_c), 
\eaen
where $y_{\max}$ is defined in Eq~(\ref{e:ymdef}), and the Sudakov form factor
\eabe S(L,\kappa,\kappa') &=& \exp\left[-\int_{\kappa'}^{\kappa}\d \kappa''2\al(\kappa'')\int_{-y_{\max}}^{y_{\max}}\d y\frac{x_1^3+x_3^3}{2}\right]  \eaen
ensures that the emitted gluon is the one with highest $k_\perp$.
An infrared cut-off $k_{\perp\mrm{cut}}$ is introduced, with $\kappa_c=\ln(k_{\perp\mrm{cut}}^2)$.
Multiplying by $S^{-1}(L,\kappa,\kappa_c)$ and taking the derivative w.r.t.\ $\kappa$ leads to, with $\gfg\equiv \gfg(L,\kappa)$, ${\gfg}_{ij}\equiv \gfg(L_{ij},\kappa)$,
\eqbe
\frac{\partial \gfg^2}{\partial \kappa} = 2\al(\kappa)\int_{-y_{\max}}^{y_{\max}}\d y\frac{x_1^3+x_3^3}2\left[{\gfg}_{12}{\gfg}_{23}{\gfg}_{13} -{\gfg}^2\right]. \label{e:gfgev}
\eqen
When $\kappa\ll L$, this has a solution linear in $L$. Parametrizing the $\kappa$ dependence as
\eqbe \ln({\gfg}(L,\kappa)) = \cAg(\kappa)+(L-\kappa-\cg)\cB(\kappa) \label{e:ABdef} \eqen
gives a pair of coupled evolution equations
\eabe
\cAg'-\cB & = & \al\exp(\cAg-\cg\cB)\sum_{n=1}\Rg_n\cB^n \label{e:cAev}, \\
\cB' & = & \al[\exp(\cAg-\cg\cB)-1]\label{e:cBev},
\eaen
where
\eqbe \Rg_n = \frac1{n!}\int_0^1\frac{\d x}x[1+(1-x)^3]\ln^n(1-x)=(-1)^n[2\zeta(n+1)-1-\frac1{2^{n+1}}-\frac1{3^{n+1}}], \label{e:Rgpdef} \eqen
$\zeta$ being the Riemann zeta-function.

\subsubsection*{\pair q Systems}
Gluon emission off a \pair q system creates a qg and a g\anti q dipole.
In addition to these,  a colour-correction term can be represented by a dipole between q and \anti q, whose emission density is weighted with $-1/\Nc^{2}$.
Following the arguments in~\cite{N3jet}, its only effect can be assumed to be a reduction of the colour factor in ``mixed'' qg dipoles from $\Nc/2$ to $\CF$, in a rapidity range $\Yq$ near the quark. For simplicity, quark masses are here neglected.

With $\gfm$ the  generating function for the multiplicity distribution in a mixed dipole, we get evolution equations in the form
\eabe
\frac{\partial \gfq}{\partial \kappa} & = & \frac{2\CF}{\Nc}\al(\kappa)\int_{-y_{\max}}^{y_{\max}}\d y\frac{x_1^2+x_3^2}2\left[{\gfm}_{12}(\Yq){\gfm}_{23}(\Yqbar)-\gfq\right] \label{e:eveqQ}, \\
\frac{\partial {\gfm}(Y)}{\partial \kappa} & = & \al(\kappa)\int_{-y_{\max}}^{y_{\max}}\d y\frac{x_1^2+x_3^3}2\left[{\gfm}_{12}(Y'){\gfg}_{23}\frac{{\gfg}_{R'}}{{\gfg}_{R}} -{\gfm}(Y)\right]\times\nonumber\\
&&\times\left[1-\frac1{\Nc^2}\Theta(Y-y_{\max}-y)\right]. \label{e:eveqM}
\eaen
Here $Y'$ represents the rapidity range of colour suppression in the quark direction, after the gluon is emitted from a mixed dipole. The original gluon of the mixed dipole is assumed to be attached to a gg dipole with logarithmic scale $L_R$, which due to recoils is reduced to $L_{R'}$. This equation contains a set of more or less free parameters, but reasonable assumptions implies
\eabe 
\ln(\gfq (L,\kappa))  & = & \cAq(\kappa) + \frac{2\CF}{\Nc}(L-\kappa-\cq)\cB(\kappa) \label{e:Aqdef}, \\
\ln({\gfm}(L,\kappa,Y)) & = & \frac12[\ln(\gfq (L,\kappa))+\ln(\gfg (L,\kappa)) +\frac1{\Nc^2}(L-2Y)\cB(\kappa)]. \eaen
This adds to the gluon dynamics result, Eq~(\ref{e:cAev}) and~(\ref{e:cBev}), a quark evolution equation
\eqbe 
\frac{\Nc}{2\CF} {\cAq}'  =  \cB + \al\exp(\cAg-\cg \cB)\sum_{n=1}\Rq_n \cB^n, \label{e:cAqev} 
\eqen
where $\Rq_n$ depend on the assumtions made on $\Yq$ and $\Yqbar$ in Eq~(\ref{e:eveqQ}). In~\cite{N3jet}, two reasonable possibilities are examined. One is $\Yq+\Yqbar=\ln(s)=L$, which implies that all $\Rq_n$ are 0. The other is $\Yq+\Yqbar=\ln(s_{\mrm{q\ol q}})=L_{13}$, which implies
\eqbe \Rq_n = \frac1{n!}(\frac{-1}{\Nc^2})^n \int_0^1\frac{\d x}x[1+(1-x)^2]\ln^n(1-x)=(\frac{1}{\Nc^2})^n[2\zeta(n+1)-1-\frac1{2^{n+1}}]. \label{e:Rqpdef} \eqen
There are some indications from experimental data in favour of this second choice~\cite{DELPHI} and we will limit ourselves to this in the present paper.

\subsubsection*{Pair Production}
To include pair production, g$\to$\pair q, (still neglecting quark masses) we add a gluon splitting term
\eqbe \frac{\Nf\al(\kappa)}{2\Nc}\int_{2k_\perp/\sqrt{s}}^1\d x_2(x_2^2+x_3^2)[{\gfm}_{12}(\Yq){\gfm}_{13}(\Yqbar)-{\gfg}^2] \eqen
to the gluon evolution equation, Eq~(\ref{e:gfgev}), and a similar term to the evolution equation for a mixed dipole, Eq~(\ref{e:eveqM}).
With reasonable assumptions about recoil effects when the gluon in a mixed dipole splits into a \pair q pair, the only modification induced by gluon splittings 
is an additional term 
\eqbe \frac{\Nf}{3\Nc}\al\left[\exp(\cAq-\cAg-[\frac{2\CF}{\Nc}\cq-\cg]\cB)\sum_{n=0}{\cB}^n\Rp_n - 1 \right]  \label{e:cApev} \eqen 
on the right-hand side of Eq~(\ref{e:cAev}). Except for $\Rp_0=1$, $\Rp_n$ depend on the assumed magnitude of the colour reduction phase space $\Yq+\Yqbar$ near the new \pair q pair. The intuitive assumption $\Yq+\Yqbar = \ln(s_{\mrm q\ol \mrm q}) = L_{23}$ implies
\eqbe \Rp_n  =  \frac3{2n!}\int_0^1\d x(x^2+(1-x)^2)\left[(1+\frac1{\Nc^2})\ln(x)+\ln(1-x)\right]^n. \label{e:Rppdef} \eqen

\subsubsection*{Combined Result}
Neglecting possible corrections related to emissions with $k_\perp$ close to the kinematical limit, discussed in section~\ref{sec:top}, the solutions to $\ln(\gf(L,\kappa))$, Eq~(\ref{e:ABdef}) and~(\ref{e:Aqdef}), are valid also when $\kappa$ is close to $L$. This implies $\cAg(L-\cg) = \ln(\gfg(L))$ and $\cAq(L-\cq)=\ln(\gfq(L))$, where $\gfg(L)$ and $\gfq(L)$ represent unbiased dipoles, i.e.\ dipoles with only kinematical constraints on $k_\perp$. Thus, Eqs~(\ref{e:cAev}),~(\ref{e:cBev}),~(\ref{e:cAqev}) and~(\ref{e:cApev}) combine to Eqs~(\ref{e:neveq}).


\begin{thebibliography}{99}
\bibitem{MLLA} \bibl{A.H. Mueller}{Nucl. Phys. }{B213}{1983}{85}\bibl{~[Erratum }{}{B241}{1984}{141}]\bibl{~}{}{B228}{1983}{351}\\
  Yu.L. Dokshitzer, S.I. Troyan, {\it in} Proc. XIX Winter School LNPI, v.I. 144 (1984); preprint LNPI-922 (1984)
\bibitem{pQCDbook}  Yu.L. Dokshitzer, V.A. Khoze, A.H. Mueller and S.I. Troyan, Basics of Perturbative QCD, ({\it Editions Fronti\`eres, Gif-sur-Yvette, 1991})
\bibitem{OPAL} 
  \bibl{OPAL collaboration}{Eur.\ Phys.\ J.}{C1}{1999}{479}
\bibitem{DELPHI} 
DELPHI collaboration, contributed paper 640 to ICHEP 2000, Osaka, Japan
\bibitem{Dremin} \bibl{A. Capella et al.}{Phys.\ Rev.}{D61}{2000}{074009}
\bibitem{Ochs} 
\bibl{S. Lupia, W. Ochs}{Phys.\ Lett.}{B388}{1996}{659};\\
\bibl{S. Lupia}{Phys.\ Lett.}{B439}{1998}{150}
\bibitem{dipolerecoil} 
\bibl{P. Ed\'en, G. Gustafson}{JHEP}{98-09}{1998}{015}
\bibitem{WWCR} 
   \bibl{G. Gustafson, U. Pettersson, P.M.Zerwas}{Phys. Lett.}{B290}{1988}{90};\\
  \bibl{V.A. Khoze, T. Sj\"ostrand}{Z. Phys.}{C62}{1994}{291};
  \bibl{}{Phys. Rev. Lett.}{72}{1994}{28};\\
  \bibl{Yu.L. Dokshitzer, V.A. Khoze, L.H. Orr, W.J. Stirling}{Nucl. Phys.}{B403}{1993}{65};\\
   \bibl{G. Gustafson, J. H\"akkinen}{Z. Phys.}{C64}{1994}{659};\\
   Report of the Working Group on ``W Mass and QCD'', hep-ph/9709283 08 Sep 1997; \\
   \bibl{J. H\"akkinen, M. Ringn\'er}{Eur.\ Phys.\ J.}{C5}{1998}{275}
\bibitem{ZCR} 
   \bibl{C. Friberg, G. Gustafson, J. H\"akkinen}{Nucl. Phys. }{B490}{1997}{289}
\bibitem{GGdipole}
  \bibl{G. Gustafson}{Nucl. Phys.}{B392} {1993} {251}
\bibitem{dipoletop} \bibl{M.Olsson, G. Gustafson}{Nucl.\ Phys.}{B406}{1993}{293}
\bibitem{MEmatch} 
  \bibl{B. Andersson, G. Gustafson, C. Sj\"ogren}{Nucl. Phys.}{B380}{1992}{391};\\
\bibl{J. Andr\'e, T. Sj\"ostrand}{Phys.\ Rev.}{D57}{1998}{5767} \\
 \bibl{S. Moretti, W.J. Stirling}{Eur.\ Phys.\ J.}{C9}{1999}{81};
\bibitem{string} 
  \bibl{B. Andersson, G. Gustafson, G. Ingelman, T. Sj\"ostrand}{Phys. Rep.}{97}{1983}{31};\\
  B. Andersson, The Lund model, ({\it Cambridge University Press, 1998}) 
\bibitem{cluster}
  \bibl{G. Marchesini et al.}{Comp. Phys. Comm.}{67}{1992}{465}
\bibitem{OPALCR} \bibl{OPAL collaboration}{Eur.\ Phys.\ J.}{C11}{1999}{217}
\bibitem{LeifCR} 
   \bibl{L. L\"onnblad}{Z. Phys.}{C70}{1996}{107}
\bibitem{N3jet} \bibl{P. Ed\'en, G. Gustafson, V. Khoze}{Eur.\ Phys.\ J.}{C11}{1999}{345}
\end{thebibliography}
\end{document}